\documentclass[11pt]{article} % use larger type; default would be 10pt

\usepackage[utf8]{inputenc} % set input encoding (not needed with XeLaTeX)
\usepackage{tikz}
\usepackage{geometry} % to change the page dimensions
\usepackage{graphicx} % support the \includegraphics command and options
\usepackage{booktabs} % for much better looking tables
\usepackage{array} % for better arrays (eg matrices) in maths
\usepackage{paralist} % very flexible & customisable lists (eg. enumerate/itemize, etc.)
\usepackage{verbatim} % adds environment for commenting out blocks of text & for better verbatim
\usepackage{subfig} % make it possible to include more than one captioned figure/table in a single float
\usepackage{multirow} % multirow in table
\usepackage{rotating} % rotate text
\usepackage[version=3]{mhchem}  % chemical formulas
\usepackage{amsmath}
\usepackage{mathabx}
\usepackage[all]{xy}
\usepackage{umoline}
\usepackage{fancyhdr} % This should be set AFTER setting up the page geometry
\usepackage{sectsty}
\allsectionsfont{\sffamily\mdseries\upshape} % (See the fntguide.pdf for font help)
\usepackage[nottoc,notlof,notlot]{tocbibind} % Put the bibliography in the ToC
\usepackage[titles,subfigure]{tocloft} % Alter the style of the Table of Contents

\title{\bf Optical selection rules in topological insulators \ce{Bi2Sb3}, \ce{Bi2Se3}, \ce{Bi2Te3} and \ce{Sb2Te3}}
\author{Jian Li,  Jiufeng J. Tu, Joseph L. Birman
\\{\em  Physics Department, The City College of New York}
\\{\em  160 Convent Avenue, New York 10031, USA}}
\date{}

\begin{document}
\maketitle
\pagenumbering{arabic}

\begin{abstract}
We performed group theoretical investigation of symmetries of excitations in topological insulators \ce{Bi2Sb3}, \ce{Bi2Te3}, \ce{Bi2Se3} and \ce{Sb2Te3}, focusing on selection rules for optical processes. Electronic transitions of bulk states to bulk states, surface states to surface states and bulk states to surface states are studied over the entire Brillouin zone. A new technique is used to deal with transitions between surface states and bulk states. Time reversal symmtry is also included in the analysis. Our results show that only \(\Gamma\), \(\Lambda\) and \(M\) points of the Brillouin zone would depend on light polarization for transitions between bulk states and surface states. As an example of application, electron spin polarizations of photoelectrons are calculated at \(\Gamma\) point. A general rule for the direct product between the representation at an arbitrary point in the Brillouin zone with itself is presented.
\end{abstract}

\begin{figure*}
\begin{center}
\includegraphics[width=12cm]{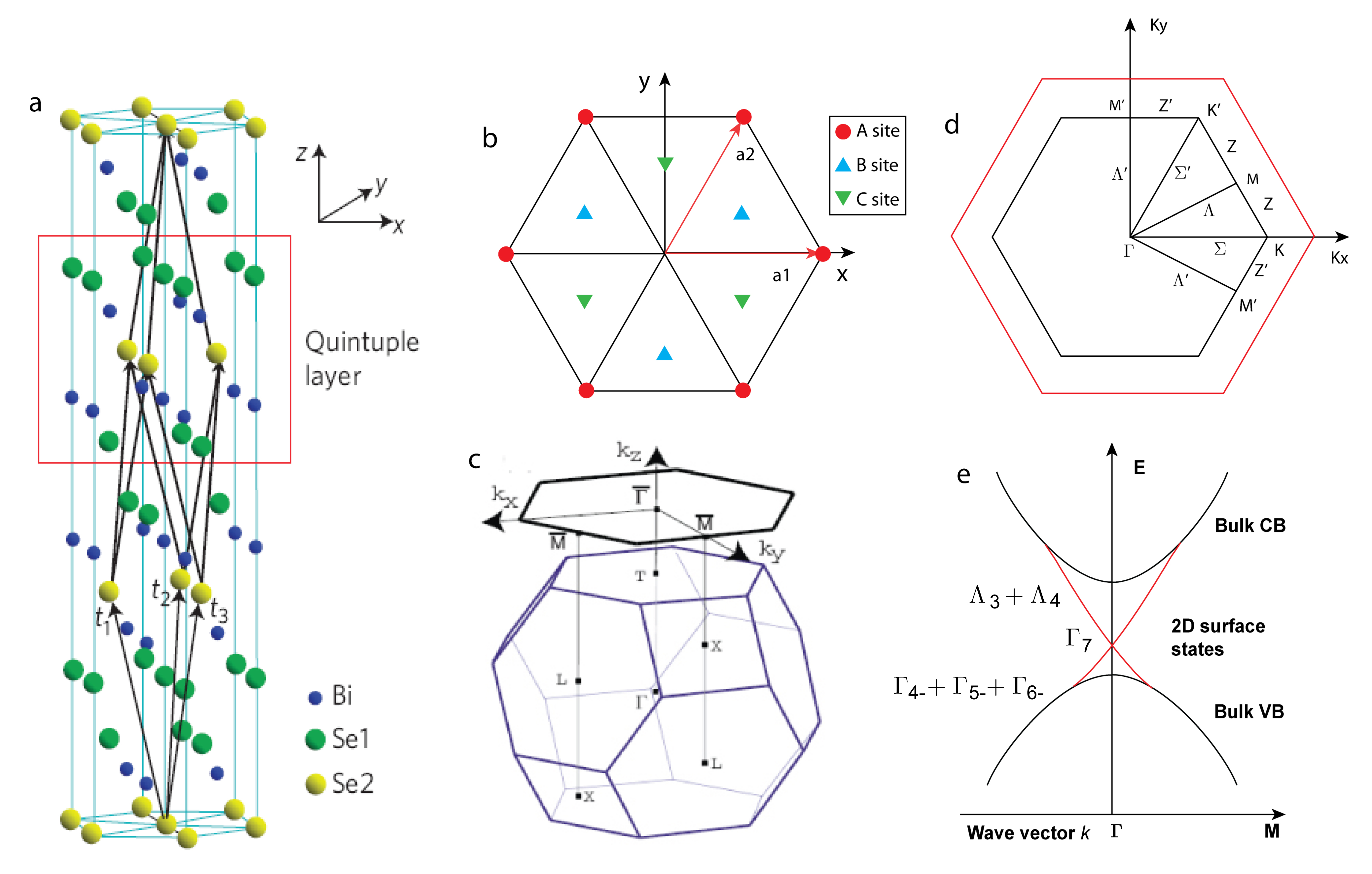}
\caption{(\(a\)): Crystal structure of \ce{Bi2Se3} from~\cite{Zhangnaturephysics}. (\(b\)): three different types of surfaces. The Brillouin zone of the bulk and the surface are shown in (\(c\)) and (\(d\)). The Brillouin zone of the surface is only part of the \(k_{z}=0\) cut of the bulk Brillouin zone (the outter hexagon in \(d\)). The labeling of the Brillouin zone points follows reference~\cite{Terzibaschian}. A schematic band structure containing both bulk states and surface states is shown in (\(e\)) along the \(\Gamma\)-\(M\) direction. The crossing point of the two surface bands has linear dispersion and is called Dirac point. The details of the band symmetry labelling are in section~\ref{sec:bandlabelling}.} \label{fig:structure}
\end{center}
\end{figure*}

{\centering {\section{INTRODUCTION}}}
Topological insulator (TI) exhibits a novel quantum state~\cite{FuPRL, MoorePRB, Bernevig, FuPRB, QiPRB}, with topologically protected conducting surface states existing within the bulk energy gap~\cite{Xia, HsiehPRL103, Chen, Zhangnaturephysics, HsiehNature, Bianchi}. In addition to their fundamental interest in quantum mechanics, such as Majorana fermions~\cite{Wilczek}, the surface states are predicted to have applications in spintronics, quantum computation and others~\cite{Hasan, Moorenature, Nayak, ZhangPRB}. The semiconducting \ce{V2VI3} compounds \ce{Bi2Sb3}, \ce{Bi2Se3}, \ce{Bi2Te3} and \ce{Sb2Te3}, formerly known as promising thermoelectric materials with figure of merit \(ZT\) = 2.5~\cite{Venkatasubramanian}, are now identified as three dimensional topological insulators. They crystallize in the layer structure (see Fig.~\ref{fig:structure}\(a\)) of rhombohedral symmetry \(R\bar{3}m\) (\(D_{3d}^{5}\))~\cite{Wyckoff}.

Compared with usual transport measurement techiques, optical measurements have the advantage that no direct contacts are required. Due to the presence of impurities, however, contribution from the surface states is masked by bulk carrier response in optical experiments~\cite{Sushkov, LaForge, Butch}. Surface sensitive tools, especially ARPES and STM, allow observations of genuine surface states. A full group theoretical study of selection rules is helpful for optical experiments because using different polarizations, contributions from the surface and the bulk might be separated. In this paper, we report optical selection rules over the entire Brillouin zone. Three kinds of electronic transitions are considered: bulk to bulk, surface to surface and bulk to surface. To apply group theory to transitions between bulk states and surface states, a smaller group consisting of all common elements from both groups is constructed. To our knowledge, this is the first time this method has been applied to bulk-surface physics problems.

This paper is arranged as follows: In section II the geometry of these TI materials is reviewed, for both bulk structure and surface structure; Section III introduces three dimensional (3D) and two dimensional (2D) space groups. Also given is the method to construct a smaller group, in which the selection rules between bulk state to surface state is calculated; Symmetries of all electronic wave functions for bulk state and surface state over the entire Brillouin zone, as well as symmetries of the surface phonon modes, are given in section IV ; Section V gives detailed selection rules at different Brillouin zone points and section VI applies the analysis to electron spin polarization calculation of photoemission process at \(\Gamma\) point. Dissusions are given in section VII, where the use of \(P6mm\) space group is justified.\\

{\centering {\section{CRYSTAL STRUCTURE AND BRILLOUIN ZONE}}}

In this section, \ce{Bi2Se3} is used to illustrate the bulk and surface structures of V-VI compounds. \ce{Bi2Sb3}, \ce{Bi2Te3} and \ce{Sb2Te3} have the same structure although the lattice parameters are different~\cite{Wyckoff}.\\

{\centering {\subsection{Crystal structure and Brillouin zone of the bulk}}}
The crystal structure of \ce{Bi2Se3} is given in Fig.~\ref{fig:structure}\(a\). The stacking of atoms in \(–z\) direction is in the order of: Se2(A) - Bi(B) - Se1(C) - Se1(A) - Bi(B) - Se2(C) - Bi(A) - Se1(B) - Se1(C) - Bi(A) - Se2(B) - Bi(C) - Se1(A) - Se1(B) - Bi(C) - Se2(A). Surface \(A\), \(B\), \(C\) sites are shown in Fig.~\ref{fig:structure}\(b\). Two neighboring Bi layers are weakly bonded by Van der Waals interaction and are most likely to produce surfaces. The crystal structure belongs to space group \(R\bar{3} m\) in Hermann-Mauguin notation and \(D_{3d}^{5}\) in Schoenflies notation (\#166). The lattice parameters are \(a\) = 4.138 \(\AA\) and \(c\) = 28.64 \(\AA\) for \ce{Bi2Se3}~\cite{Wyckoff}. Fig.~\ref{fig:structure}\(a\) also shows the Bravias lattice vectors:
\begin{eqnarray}
\stackrel{\rightharpoonup}{{a}_{1}} & = & (\frac{-a}{2}, \frac{-a}{2\sqrt{3}}, \frac{c}{3});\nonumber \\
\stackrel{\rightharpoonup}{{a}_{2}} & = & (\frac{a}{2}, \frac{-a}{2\sqrt{3}}, \frac{c}{3});\nonumber \\
\stackrel{\rightharpoonup}{{a}_{3}} & = & (0, \frac{-a}{\sqrt{3}}, \frac{c}{3}). \nonumber
\end{eqnarray}
This is the rhombohedral translational group {\em R}. The hexagonal unit cell contains nine formula units and the primitive cell contains one formula unit. Two Bi atoms and two Se1 atoms occupy \(2c\) Wyckoff positions and one Se2 atom occupies \(1a\) Wyckoff position~\cite{Wyckoff}. The corresponding reciprocal lattice vectors are:
\begin{eqnarray}
\stackrel{\rightharpoonup}{{b}_{1}} & = & (\frac{-2\pi}{a}, \frac{-2\pi}{\sqrt{3}a}, \frac{2\pi}{c});\nonumber \\
\stackrel{\rightharpoonup}{{b}_{2}} & = & (\frac{2\pi}{a}, \frac{-2\pi}{\sqrt{3}a}, \frac{2\pi}{c});\nonumber \\
\stackrel{\rightharpoonup}{{b}_{3}} & = & (0, \frac{-4\pi}{\sqrt{3}a}, \frac{2\pi}{c}).\nonumber 
\end{eqnarray}
The first Brillouin zone is shown in Fig.~\ref{fig:structure}\(c\). High symmetry points on the surface are labelled.\\

{\centering {\subsection{Crystal structure and Brillouin zone of the surface}}}

Surfaces are most likely to form on the (111) plane, between the layers. There are three types of possible surface structures, labeled \(A\), \(B\) and \(C\) in Fig.~\ref{fig:structure}\(b\). \(B\) and \(C\) differ from \(A\) by a fractional translation (\({{{a}/{2}}, \pm{{a}/{2\sqrt{3}}}}, \pm{\frac{c}{15}}\)). All three surface structures belong to ``wallpaper'' group \(P6mm\) (see section~\ref{sec:spacegroup}\ref{sec:2dspacegroup} for detail). The lattice parameter is \(a\) = 4.138 \(\AA\). Fig.~\ref{fig:structure}\(b\) also shows the Bravias lattice vectors of the surface structure:

\begin{eqnarray}
\stackrel{\rightharpoonup}{{a'}_{1}} & = & (a, 0, 0);\nonumber \\
\stackrel{\rightharpoonup}{{a'}_{2}} & = & (\frac{a}{2}, \frac{-\sqrt{3}a}{2},0). \nonumber
\end{eqnarray}
They form translational group {\em P} and can be decomposed using the bulk Bravais lattice:
\begin{eqnarray}
\stackrel{\rightharpoonup}{{a'}_{1}}&=&\stackrel{\rightharpoonup}{a}_{2} - \stackrel{\rightharpoonup}{a}_{1}; \nonumber \\
\stackrel{\rightharpoonup}{{a'}_{2}} & = &\stackrel{\rightharpoonup}{a}_{3} - \stackrel{\rightharpoonup}{a}_{1}. \nonumber
\end{eqnarray}
For all \(A\), \(B\) and \(C\) surfaces, there is only one atom in one primitive cell. It occupies Wyckoff position \(1(a)\)~\cite{Wyckoff}. The reciprocal lattice vectors are:
\begin{eqnarray}
\stackrel{\rightharpoonup}{{b'}_{1}} & = & (\frac{2\pi}{a}, \frac{-2\pi}{\sqrt{3}a}, 0);\nonumber \\
\stackrel{\rightharpoonup}{{b'}_{2}} & = & (0, \frac{4\pi}{\sqrt{3}a}, 0).\nonumber
\end{eqnarray}
The first Brillouin zone of the surface is shown in Fig.~\ref{fig:structure}\(d\), with high symmetry points labeled \(\Gamma\), \(M\), \(K\), high symmetry lines labeled \(\Lambda\), \(\Sigma\), \(Z\) and arbitrary point \(R\). Brillouin zone labelings are adopted from reference~\cite{Terzibaschian}. The Brillouin zone of \(P6mm\) is only a part of the \(k_{z}=0\) cut of the Brillouin zone of \(R\bar{3}m\) (Fig.~\ref{fig:structure}\(d\)). The zone boudary points of \(P6mm\) is still inside the first Brillouin zone of \(R\bar{3}m\).\\

{\centering {\section{SPACE GROUPS} \label{sec:spacegroup}}}

Group theory deals with symmetries. Its applications in physics problems are based on the matrix-element theorem: 
\begin{equation}
\int \psi \hat{H} \phi \text{ d}\tau \neq 0 \label{eqn:matrixelementtheorem}
\end{equation}
only if \(\Gamma_{1} \in {D^{\psi}}\otimes{D^{H}}\otimes{D^{\phi}}\). \(\psi\), \(\hat{H}\), \(\phi\) are wave functions and operators; \(D^{\psi}\), \(D^{H}\), \(D^{\phi}\) are the representations according to which \(\psi\), \(\hat{H}\), \(\phi\) transform; \(\otimes\) is the direct product and \(\Gamma_{1}\) is the identity representation. This theorem leads to well known selection rules in infrared absorption, Raman scattering as well as many others~\cite{Tinkham}. To avoid confusions, we use \(D_{i}\), \(d_{i}\) and \(\delta_{i}\) to denote the \(i\)th representation of space group \(R\bar{3}m\), \(P6mm\) and \(P3m1\), respectively. \(P3m1\) is a subgroup of both \(R\bar{3}m\) and \(P6mm\):
\begin{displaymath}
\xymatrix{
{\framebox[1.5cm][c]{\(R\bar{3}m\)}} \ar[ddr]& & {\framebox[1.5cm][c]{\(P6mm\)}} \ar[ddl] \\
&&\\
&{\framebox[1.5cm][c]{\(P3m1\)}}& }
\end{displaymath}\\

{\centering {\subsection{Three dimensional space group}}} 

The crystal structure of \ce{Bi2Se3} belongs to space group \(R\bar{3}m\) (\(D_{3d}^{5}\), \#166). The symmetry operations are~\cite{Hahn}: \(\{E|0\}\), \(\{C_{6}^{2}|0\}\), \(\{C_{6}^{4}|0\}\), \(\{C_{x}|0\}\), \(\{C_{6}^{2}C_{x}|0\}\), \(\{C_{6}^{4}C_{x}|0\}\), \(\{I|0\}\), \(\{IC_{6}^{2}|0\}\), \(\{IC_{6}^{4}|0\}\), \(\{IC_{x}|0\}\), \(\{IC_{6}^{2}C_{x}|0\}\), \(\{IC_{6}^{4}C_{x}|0\}\), as well as their products with translational group \(R\). All symmetry operations have their usual meanings~\cite{Koster}: \(E\) is the identity operation, \(C_{m}^{n}\) is the rotation counterclockwise by \({2\pi \cdot n}/{m}\), \(C_{x}\) is the rotation along \(x\) axis by \(\pi\) and \(I\) is the inversion. Symmetry elements are expressed in terms of generators for better presentation. Operations \(C_{3}^{i}\) are written down as \(C_{6}^{2i}\), which makes connections to two dimensional space groups more clear. Space group \(R\bar{3}m\) is symmorphic. The character table of irreducible representations at any point of the Brillouin zone can be easily obtained from their corresponding point groups~\cite{Koster}. When optical transitions connecting two bulk electronic states are considered, both electronic wave functions \(\psi\) and \(\phi\) of equation~(\ref{eqn:matrixelementtheorem}) transform according to certain irreducible representations of \(R\bar{3}m\). Direct product within \(R\bar{3}m\) provides selection rules.\\

{\centering {\subsection{Two dimensional space group} \label{sec:2dspacegroup}}} 

Symmetries of surfaces can be described by two dimensional space groups, which are also called ``wallpaper group'', ``plane symmetry group'' or ``plane crystallographic group''. There are totally 17 wallpaper groups, catagorized into 4 crystal systems ({\em oblique}, {\em rectangular}, {\em square}, {\em hexagonal}) and 5 Bravais classes ({\em p-oblique}, {\em p-rectangular}, {\em c-rectangular}, {\em p-square}, {\em p-hexagonal}), just as 230 3D space groups belong to 7 crystal systems and 14 Bravais classes.

The structures of type \(A\), \(B\) and \(C\) surfaces are the same (see Fig.~\ref{fig:structure}\(b\)). The Bravais lattice class is {\em p-hexagonal} and the crystal system is {\em hexagonal}~\cite{Terzibaschian}. Symmetry operations are~\cite{Hahn}: \(\{E|0\}\), \(\{C_{6}|0\}\), \(\{C_{6}^{2}|0\}\), \(\{C_{6}^{3}|0\}\), \(\{C_{6}^{4}|0\}\), \(\{C_{6}^{5}|0\}\),\(\{IC_{x}|0\}\), \(\{IC_{6}C_{x}|0\}\), \(\{IC_{6}^{2}C_{x}|0\}\), \(\{IC_{6}^{3}C_{x}|0\}\), \(\{IC_{6}^{4}C_{x}|0\}\), \(\{IC_{6}^{5}C_{x}|0\}\), as well as their products with translational group \(P\). This is \(P6mm\) wallpaper group (\#17). \(P6mm\) is symmorphic and point group character tables are sufficient for all irreducible representations in the entire Brillouin zone. Both \(\psi\) and \(\phi\) of equation~(\ref{eqn:matrixelementtheorem}) transform according to certain irreducible representations of \(P6mm\). Direct products within \(P6mm\) give all the selection rules. We abandoned the barred notations for surface states and it should not be confusing.\\

{\centering {\subsection{Transition between 3D and 2D}}}

When optical transitions happen between a bulk electronic state and a surface electronic state, the above calculations need to be modified. Let \(\psi\) denote the bulk electronic state that transforms as \(D_{i}\) of \(R\bar{3}m\) and \(\phi\) denote the surface electronic state that transforms as \(d_{j}\) of \(P6mm\). The matrix-element theorem does not apply to \(D_{i}\) and \(d_{j}\) because they are representations of different groups.  In order to apply group theory, a smaller group \(G'\) must be constructed, containing all common symmetry operations from both \(R\bar{3}m\) and \(P6mm\). In the smaller group \(G'\),  \(D_{i}\) of \(R\bar{3}m\) restricts to irreducible representation \(\delta_{i}\) of \(G'\) and \(d_{j}\) of \(P6mm\) restricts to irreducible representation \(\delta_{j}\) of \(G'\). Matrix-element theorem can now be applied since \(\delta_{i}\) and \(\delta_{j}\) are both irreducible representations of \(G'\). 

The common symmetry operations of \(R\bar{3}m\) and \(P6mm\) are: \(\{E|0\}\), \(\{C_{6}^{2}|0\}\), \(\{C_{6}^{4}|0\}\), \(\{IC_{x}|0\}\), \(\{IC_{6}^{2}C_{x}|0\}\), \(\{IC_{6}^{4}C_{x}|0\}\), and their products with translational group \(P\). This is the wallpaper group \(P3m1\) (\#14)~\cite{Terzibaschian} and it belongs to {\em hexagonal} crystal system and {\em p-hexagonal} Bravais class. It is also symmorphic. The same translational group \(P\) as in \(P6mm\) leads to same Brillouin zone. However, they have different point groups: \(C_{3v}\)(\(3m\)) for \(P3m1\) and \(C_{6v}\)(\(6mm\)) for \(P6mm\). This is important in later calculations. For example, \(\Sigma\) point and \(\Sigma'\) point in Fig.~\ref{fig:structure}\(d\) are equivalent in \(P6mm\) while in \(P3m1\) they are inequivalent.\\

{\centering {\section{ELECTRONIC WAVE FUNCTIONS and SURFACE PHONONS}}}

{\centering {\subsection{Symmetries of electronic wave functions \label{sec:bandlabelling}}}}

In the tight binding approximation, the electronic wave functions are formed by atomic orbitals localized at individual atoms. The symmetry of the wave functions is the direct product of the permutation representation and the atomic orbital representation. The permutation representation takes atomic positions as basis functions and its characters are the number of atoms unchanged or can be shifted back to itself through a lattice vector. The atomic orbital representation is simply the representation formed by basis functions of atomic orbitals: (\(x^{2}+y^{2}+z^{2}\)) for \(s\), (\(x, y, z\)) for \(p\) and so on.

The electronic wave function symmetries for the bulk are given in Table~\ref{tab:electronicbandd3d}. The five atoms in \(R\bar{3}m\) occupy two Wyckoff positions \(2(c)\) and \(1(a)\). Wave functions are listed for atomic orbitals \(s\), \(p\), \(d\), \(f\), which include all atomic orbitals in \ce{Bi}, \ce{Se},  \ce{Sb} and \ce{Te}. Table~\ref{tab:electronicbandp6mm} lists possible electronic wave function symmetries for the surface where atoms only occupy \(1(a)\) Wyckoff position. Wave fucntions are also listed for atomic orbitals \(s\), \(p\), \(d\), \(f\). The surface states in \ce{Bi2Se3} are formed by atomic orbitals \(p_{z}\)~\cite{Zhangnaturephysics} and they belong to representation \(\Gamma_{1}\), \(\Lambda_{2}\), \(\Sigma_{2}\), \(M_{1}\), \(K_{1}\), \(Z_{2}\) and \(R_{1}\) in the surface space group \(P6mm\).The lowest conduction band and highest valence band of bulk electronic states are formed by atomic orbits \(P_{x}, P_{y}\)~\cite{Zhangnaturephysics}. Taking into account of electron spins, the electronic bands  along the \(\Gamma M\) line in Fig.~\ref{fig:structure}\(e\) should be labelled \(\Gamma_{7}\) and \(\Lambda_{3} \oplus \Lambda_{4}\) for the surface states and \(\Gamma_{4-}\), \(\Gamma_{5-} \oplus \Gamma_{6-}\) and \(\Lambda_{3} \oplus \Lambda_{4}\) for the bulk states.\\

\begin{table*}
\begin{center}
\begin{tabular}{ccccccccc} \hline
positions&orbitals& \(\Gamma\) & \(\Lambda\)& \(\Sigma\)& \(M\)& \(K\)& \(Z\)& \(R\)  \\ \hline
\(2(c)\) & \(s\) & \(\Gamma_{1+}\)\(\Gamma_{2-}\)  & \(\Lambda_{1}\) &\(\Sigma_{1}\)\(\Sigma_{2}\)&\(M_{1}\)&\(K_{1}\)\(K_{2}\)&\(Z_{1}\)&\(R_{1}\) \\
& \(p\)& \(\Gamma_{1+}\)\(\Gamma_{3+}\)\(\Gamma_{2-}\)\(\Gamma_{3-}\)& \(\Lambda_{1}\)\(\Lambda_{2}\)&\(\Sigma_{1}\)\(\Sigma_{2}\)&\(M_{1}\)\(M_{2}\)&\(K_{1}\)\(K_{2}\)&\(Z_{1}\)&\(R_{1}\) \\
& \(d\)& \(\Gamma_{1+}\)\(\Gamma_{3+}\)\(\Gamma_{2-}\)\(\Gamma_{3-}\)& \(\Lambda_{1}\)\(\Lambda_{2}\)&\(\Sigma_{1}\)\(\Sigma_{2}\)&\(M_{1}\)\(M_{2}\)&\(K_{1}\)\(K_{2}\)&\(Z_{1}\)&\(R_{1}\) \\
& \(f\)& \(\Gamma_{1+}\)\(\Gamma_{2+}\)\(\Gamma_{3+}\)\(\Gamma_{1-}\)\(\Gamma_{2-}\)\(\Gamma_{3-}\)& \(\Lambda_{1}\)\(\Lambda_{2}\) &\(\Sigma_{1}\)\(\Sigma_{2}\)&\(M_{1}\)\(M_{2}\)&\(K_{1}\)\(K_{2}\)&\(Z_{1}\)&\(R_{1}\) \\ \hline
\(1(a)\) & \(s\) & \(\Gamma_{1+}\) & \(\Lambda_{1}\)&\(\Sigma_{1}\)&\(M_{1}\)&\(K_{1}\)&\(Z_{1}\)&\(R_{1}\) \\
& \(p\)& \(\Gamma_{2-}\)\(\Gamma_{3-}\)& \(\Lambda_{1}\)\(\Lambda_{2}\) &\(\Sigma_{1}\)\(\Sigma_{2}\)&\(M_{1}\)\(M_{2}\)&\(K_{1}\)\(K_{2}\)&\(Z_{1}\)&\(R_{1}\) \\
& \(d\)& \(\Gamma_{1+}\)\(\Gamma_{3+}\) & \(\Lambda_{1}\)\(\Lambda_{2}\) &\(\Sigma_{1}\)\(\Sigma_{2}\)&\(M_{1}\)\(M_{2}\)&\(K_{1}\)\(K_{2}\)&\(Z_{1}\)&\(R_{1}\) \\
& \(f\)& \(\Gamma_{1-}\)\(\Gamma_{2-}\)\(\Gamma_{3-}\)& \(\Lambda_{1}\)\(\Lambda_{2}\) &\(\Sigma_{1}\)\(\Sigma_{2}\)&\(M_{1}\)\(M_{2}\)&\(K_{1}\)\(K_{2}\)&\(Z_{1}\)&\(R_{1}\) \\ \hline
\end{tabular}
\caption{Electronic wave function symmetry in \(R\bar{3}m\).} \label{tab:electronicbandd3d}
\end{center}
\end{table*}

\begin{table*}
\begin{center}
\begin{tabular}{ccccccccc} \hline
position&orbitals& \(\Gamma\) & \(\Lambda\)& \(\Sigma\)& \(M\)& \(K\)& \(Z\)& \(R\)  \\ \hline
\(1(a)\) & \(s\) & \(\Gamma_{1}\) & \(\Lambda_{1}\)&\(\Sigma_{1}\)&\(M_{1}\)&\(K_{1}\)&\(Z_{1}\)&\(R_{1}\) \\
& \(p\)& \(\Gamma_{1}\)\(\Gamma_{5}\)& \(\Lambda_{1}\)\(\Lambda_{2}\) &\(\Sigma_{1}\)\(\Sigma_{2}\)&\(M_{1}\)\(M_{3}\)\(M_{4}\)&\(K_{1}\)\(K_{3}\)&\(Z_{1}\)\(Z_{2}\)&\(R_{1}\) \\
& \(d\)& \(\Gamma_{1}\)\(\Gamma_{5}\)\(\Gamma_{6}\) & \(\Lambda_{1}\)\(\Lambda_{2}\) &\(\Sigma_{1}\)\(\Sigma_{2}\)&\(M_{1}\)\(M_{2}\)\(M_{3}\)\(M_{4}\)&\(K_{1}\)\(K_{3}\)&\(Z_{1}\)\(Z_{2}\)&\(R_{1}\) \\
& \(f\)&  \(\Gamma_{1}\)\(\Gamma_{3}\)\(\Gamma_{4}\)\(\Gamma_{5}\)\(\Gamma_{6}\)& \(\Lambda_{1}\)\(\Lambda_{2}\) &\(\Sigma_{1}\)\(\Sigma_{2}\)&\(M_{1}\)\(M_{2}\)\(M_{3}\)\(M_{4}\)&\(K_{1}\)\(K_{2}\)\(K_{3}\)&\(Z_{1}\)\(Z_{2}\)&\(R_{1}\) \\ \hline
\end{tabular}
\caption{Electronic wave function symmetry in \(P6mm\).} \label{tab:electronicbandp6mm}
\end{center}
\end{table*}

{\centering {\subsection{Symmetries of surface phonon modes}}}

For the sake of completeness, we list also the surface phonon structures of TIs (the phonon structure of the bulk \ce{Bi2Se3} is well known~\cite{Richter}). With the assumption that the surface modes are strongly localized and symmetry classification of surface phonons can be carried out within \(P6mm\) space group. \(\Gamma_{1}\) and \(\Gamma_{6}\) phonon modes are Raman active in \(C_{6v}\) point group. The corresponding two dimensional Raman tensors are: \(\begin{pmatrix}
  a & 0 \\
  0 & a 
 \end{pmatrix}\) for \(\Gamma_{1}\) and \(\begin{pmatrix}
  b & 0 \\
  0 & -b 
 \end{pmatrix}\), \(\begin{pmatrix}
  0 & b \\
  b & 0 
 \end{pmatrix}\) for \(\Gamma_{6}\).\\

\begin{table*}
\begin{centering}
\begin{tabular}{cccccccc} \hline
&\(\Gamma\) & \(\Lambda\) & \(\Sigma\) & \(M\) & \(K\) & \(Z\) & \(R\) \\ \hline
\(x, y\) & \(\Gamma_{5}\) & 2\(\Lambda_{1}\) & 2\(\Sigma_{1}\) & \(M_{3}+M_{4}\) & \(K_{3}\) & 2\(Z_{1}\) & 2\(R_{1}\) \\
\(z\) & \(\Gamma_{1}\) & \(\Lambda_{2}\) & \(\Sigma_{2}\) & \(M_{1}\) & \(K_{1}\) & \(Z_{2}\) & \(R_{1}\) \\ \hline
\end{tabular}
\caption{Surface phonon modes of \ce{Bi2Se3} at different Brillouin zone points. \(x, y\) row gives atomic displacements confined in the surface while \(z\) row gives atomic displacements out of the surface.} \label{tab:surfacephonon}
\end{centering}
\end{table*}

{\centering {\section{OPTICAL SELECTION RULES}}}

In this section, all optical selection rules at different Brillouin points are tabulated. In each Brillouin zone point, the group of wave vector \(G_{0}{(k)}\) and factor group \(G/{G_{0}{(k)}}\) are listed. The tables for optical transition selection rules are obtained by performing direct products within \(R\bar{3}m\) and \(P6mm\), as well as direct products between \(R\bar{3}m\) and \(P6mm\). Restrictions from \(R\bar{3}m\) and \(P6mm\) at different Brillouin zone points are given in Table~\ref{tab:restriction}. Only zone center representations are listed in the direct product because only those participate in single photon optical transitions. For bulk electronic bands, only states lies in the \(k_{z}=0\) plane can be optically excited to surface electronic states.\\

{\centering {\subsection{Direct products for space groups}}}

The direct product of two irreducible representations \(D^{(\bigstar{K})(m)}\) \(\otimes\) \(D^{(\bigstar{K'})(m')}\) is in general reducible and the reduction coefficients \((\bigstar{k}m\bigstar{k'}m'|\bigstar{k''}m'')\) can be determined once the irreducible representaion \(D^{(\bigstar{K})(m)}\) itself is obtained~\cite{Birman}. Rotational operations bring wave functions with wave vector \(k\) to wave functions with wave vectors \(\{\varphi \cdot k\}\). Some of these wave vectors are equivalent to \(k\): \(\{\varphi_{i} \cdot k\}\) = \(k\) + \(B_{H}\), where \(B_{H}\) is any reciprocal lattice vector. The inequivalent set of \(\{\varphi \cdot k\}\) is defined as {\em star of \(k\)} and each inequivalent one of \(\{\varphi \cdot k\}\) is called an {\em arm} of the star. Those symmetry operations \(\{\varphi|t(\varphi)\}\) that \(\{\varphi_{i} \cdot k\}\) = \(k\) + \(B_{H}\) defines a new group \({G}_{0}(k)\). It is called {\em group of wave vector \(k\)} and is a subgroup of \(G\). Irreducible representations of space group \(G\) can be induced from irreducible representations of space group \({G}_{0}(k)\): \(D^{({k})(m)}\)\(\uparrow\)\(D^{(\bigstar{k})(m)}\).

Time reversal operation \(\hat \theta\) is also a symmetry operation of the system. Taking into account the time reversal operator \(\hat \theta\), a larger group \(H = G \oplus \hat{\theta}G\) is obtained. \(H\) is the (second type) magnetic space group and \(G\) is the original space group of only spatial operations. The represenation of \(H\) can be induced from \(G\) and this problem was treated by Dimmock and Wheeler~\cite{Dimmock}. It can be shown~\cite{Birman} that if \(-k\) \(\notin\) \(\bigstar{k}\), there is extra degeneracy that representations from \(\bigstar{-k}\) and \(\bigstar{k}\) are forced to stick together. When \(-k\) \(\in\) \(\bigstar{k}\), there are still possibilities that time reversal operation connects two irreducible representations, such as \(\Gamma_{2}\) and \(\Gamma_{3}\) of \(C_{3v}\). Those representations must appear in pairs, unless magnetic field breaks the time reversal symmetry. Representations from its time reversal star \(\bigstar{-k}\) are labelled as \(D^{*}\), such as \(\Lambda_{1}\) and \(\Lambda^{*}_{1}\) in table~\ref{tab:restriction} are representations at \(\bigstar{\Lambda}\) and \(\bigstar{-\Lambda}\) of \(P3m1\).\\

\begin{table*}
\begin{center}
\begin{tabular}{cccc} \hline
\(R\bar{3}m\) (\(D_{i}\))& \(P3m1\) (\(\delta_{i}\))& \(P6mm\) (\(d_{i}\))& \(P3m1\) (\(\delta_{i}\)) \\ \hline
\(\Gamma_{1+}\) & \(\Gamma_{1}\) &\(\Gamma_{1}\) &\(\Gamma_{1}\) \\ 
\(\Gamma_{2+}\) & \(\Gamma_{2}\) &\(\Gamma_{2}\) &\(\Gamma_{2}\) \\ 
\(\Gamma_{3+}\) & \(\Gamma_{3}\) &\(\Gamma_{3}\) &\(\Gamma_{2}\) \\ 
\(\Gamma_{1-}\) & \(\Gamma_{2}\) &\(\Gamma_{4}\) &\(\Gamma_{1}\) \\ 
\(\Gamma_{2-}\) & \(\Gamma_{1}\) &\(\Gamma_{5}\) &\(\Gamma_{3}\) \\ 
\(\Gamma_{3-}\) & \(\Gamma_{3}\) &\(\Gamma_{6}\) &\(\Gamma_{3}\) \\ \hline
\(\Lambda_{1}\) & \(\Lambda_{1} \oplus \Lambda_{1}^{*} \) & \(\Lambda_{1}\) & \(\Lambda_{1} \oplus \Lambda_{1}^{*} \) \\
\(\Lambda_{2}\) & \(\Lambda_{2} \oplus \Lambda_{2}^{*} \) & \(\Lambda_{2}\) & \(\Lambda_{2} \oplus \Lambda_{2}^{*} \) \\ \hline
\(\Sigma_{1}\) & \(\Sigma_{1}\) & \(\Sigma_{1}\) & \(\Sigma_{1}\) \\
\(\Sigma_{2}\) & \(\Sigma_{2}\) & \(\Sigma_{2}\) & \(\Sigma_{2}\) \\ \hline
\(M_{1}\) & \(M_{1}\) & \(M_{1}\) & \(M_{1}\) \\
\(M_{2}\) & \(M_{2}\) & \(M_{2}\) & \(M_{2}\) \\
&& \(M_{3}\) & \(M_{2}\) \\
&& \(M_{4}\) & \(M_{1}\) \\ \hline
\(K_{1}\) & \(K_{1}+ K_{2} \oplus K_{3}\) & \(K_{1}\) & \(K_{1}\) \\
\(K_{2}\) & \(K_{1}+ K_{2} \oplus K_{3}\) & \(K_{2}\) & \(K_{1}\) \\
&& \(K_{3}\) & \(K_{2}+K_{3}\) \\ \hline
\(Z_{1}\) & \(Z_{1}\oplus Z_{1}^{*}\) & \(Z_{1}\) & \(Z_{1}\oplus Z_{1}^{*}\) \\
&& \(Z_{2}\) & \(Z_{1}\oplus Z_{1}^{*}\) \\ \hline
\(R_{1}\) & \(R_{1}\oplus R_{1}^{*}\) & \(R_{1}\) & \(R_{1}\oplus R_{1}^{*}\) \\ \hline 
\end{tabular}
\caption{Restrictions from \(R\bar{3}m\) and \(P6mm\) to \(P3m1\). The * means representations from its time reversal star. \(K_{2}\) and \(K_{3}\) are also connected by time reversal symmetry, although they belong to the same star \(\bigstar K\).} \label{tab:restriction}
\end{center}
\end{table*}

{\centering {\subsection{\(\Gamma\) point}}}

\(\Gamma\) = (0, 0). In \(R\bar{3}m\),  \(G_{0}{(k)}\)= (\(\{E|0\}\), \(\{C_{6}^{2}|0\}\), \(\{C_{6}^{4}|0\}\), \(\{C_{x}|0\}\), \(\{C_{6}^{2}C_{x}|0\}\), \(\{C_{6}^{4}C_{x}|0\}\), \(\{I|0\}\), \(\{IC_{6}^{2}|0\}\), \(\{IC_{6}^{4}|0\}\), \(\{IC_{x}|0\}\), \(\{IC_{6}^{2}C_{x}|0\}\), \(\{IC_{6}^{4}C_{x}|0\}\)) = \(D_{3d}\), \(G/{G_{0}{(k)}}\)=(\(\{E|0\}\)). In \(P6mm\), \(G_{0}{(k)}\) = (\(\{E|0\}\), \(\{C_{6}|0\}\), \(\{C_{6}^{2}|0\}\), \(\{C_{6}^{3}|0\}\), \(\{C_{6}^{4}|0\}\), \(\{C_{6}^{5}|0\}\),\(\{IC_{x}|0\}\), \(\{IC_{6}^{1}C_{x}|0\}\), \(\{IC_{6}^{2}C_{x}|0\}\), \(\{IC_{6}^{3}C_{x}|0\}\), \(\{IC_{6}^{4}C_{x}|0\}\), \(\{IC_{6}^{5}C_{x}|0\}\)) = \(C_{6v}\), \(G/{G_{0}{(k)}}\) = (\(\{E|0\}\)). In \(P3m1\), \(G_{0}{(k)}\) = (\(\{E|0\}\), \(\{C_{6}^{2}|0\}\), \(\{C_{6}^{4}|0\}\), \(\{IC_{x}|0\}\), \(\{IC_{6}^{2}C_{x}|0\}\), \(\{IC_{6}^{4}C_{x}|0\}\)) = \(C_{3v}\), \(G/{G_{0}{(k)}}\)=(\(\{E|0\}\)).\\

\begin{table*} \footnotesize
\begin{center}
\begin{tabular}{cccccccccccccc} \hline
&& \multicolumn{6}{c}{\(R\bar{3}m\)}& \multicolumn{6}{c}{\(P6mm\)} \\ \hline
&& \(\Gamma_{1+}\) & \(\Gamma_{2+}\) & \(\Gamma_{3+}\) & \(\Gamma_{1-}\)& \(\Gamma_{2-}\) & \(\Gamma_{3-}\) &\(\Gamma_{1}\) & \(\Gamma_{2}\) & \(\Gamma_{3}\) & \(\Gamma_{4}\) & \(\Gamma_{5}\) & \(\Gamma_{6}\)\\ \hline
\multirow{6}{*}{\rotatebox{270}{\mbox{\(\tiny{R\bar{3}m}\)}}} & \(\Gamma_{1+}\) &\(\Gamma_{1}\)  & \(\Gamma_{2}\)  & \(\Gamma_{3}\)  & \(\Gamma_{4}\)  & \(\Gamma_{5}\)  & \(\Gamma_{6}\)  &\(\Gamma_{1}\)  & \(\Gamma_{2}\)  & \(\Gamma_{2}\)  & \(\Gamma_{1}\)  & \(\Gamma_{3}\)  & \(\Gamma_{3}\)  \\
 & \(\Gamma_{2+}\) && \(\Gamma_{2}\)  & \(\Gamma_{3}\)  & \(\Gamma_{4}\)  & \(\Gamma_{5}\)  & \(\Gamma_{6}\)  &\(\Gamma_{2}\)  & \(\Gamma_{1}\)  & \(\Gamma_{1}\)  & \(\Gamma_{2}\)  & \(\Gamma_{3}\)  & \(\Gamma_{3}\)  \\
& \(\Gamma_{3+}\) &&& \(\Gamma_{1}\)\(\Gamma_{2}\)\(\Gamma_{3}\)   & \(\Gamma_{6}\)  & \(\Gamma_{6}\)  & \(\Gamma_{4}\)\(\Gamma_{5}\)\(\Gamma_{6}\)  &\(\Gamma_{3}\)  & \(\Gamma_{3}\)  & \(\Gamma_{3}\)  & \(\Gamma_{3}\)  & \(\Gamma_{1}\)\(\Gamma_{2}\)\(\Gamma_{3}\)  & \(\Gamma_{1}\)\(\Gamma_{2}\)\(\Gamma_{3}\)  \\
& \(\Gamma_{1-}\) &&&& \(\Gamma_{1}\) & \(\Gamma_{2}\) & \(\Gamma_{3}\) &\(\Gamma_{2}\) & \(\Gamma_{1}\) & \(\Gamma_{1}\) & \(\Gamma_{2}\) & \(\Gamma_{3}\) & \(\Gamma_{3}\) \\
& \(\Gamma_{2-}\) &&&&&\(\Gamma_{1}\)& \(\Gamma_{3}\) &\(\Gamma_{1}\) & \(\Gamma_{2}\) & \(\Gamma_{2}\) & \(\Gamma_{1}\) & \(\Gamma_{3}\) & \(\Gamma_{3}\) \\
& \(\Gamma_{3-}\) &&&&&& \(\Gamma_{1}\)\(\Gamma_{2}\)\(\Gamma_{3}\) &\(\Gamma_{3}\) & \(\Gamma_{3}\) & \(\Gamma_{3}\) & \(\Gamma_{3}\) & \(\Gamma_{1}\)\(\Gamma_{2}\)\(\Gamma_{3}\) & \(\Gamma_{1}\)\(\Gamma_{2}\)\(\Gamma_{3}\) \\ \hline
\multirow{6}{*}{\rotatebox{270}{\mbox{\(P6mm\)}}} & \(\Gamma_{1}\)  & & & & & & &\(\Gamma_{1}\) & \(\Gamma_{2}\) & \(\Gamma_{3}\)  & \(\Gamma_{4}\)  & \(\Gamma_{5}\)  & \(\Gamma_{6}\)  \\
 & \(\Gamma_{2}\) & & & & & & && \(\Gamma_{1}\)  & \(\Gamma_{4}\)  & \(\Gamma_{3}\)  & \(\Gamma_{5}\)  & \(\Gamma_{6}\)  \\
& \(\Gamma_{3}\) & & & & & & && & \(\Gamma_{1}\)  & \(\Gamma_{2}\)  & \(\Gamma_{6}\)  & \(\Gamma_{5}\)  \\
& \(\Gamma_{4}\) & & & & & & && & & \(\Gamma_{1}\)  & \(\Gamma_{6}\)  & \(\Gamma_{5}\)  \\
& \(\Gamma_{5}\) & & & & & & && & & & \(\Gamma_{1}\)\(\Gamma_{2}\)\(\Gamma_{6}\) & \(\Gamma_{3}\)\(\Gamma_{4}\)\(\Gamma_{5}\) \\
& \(\Gamma_{6}\) & & & & & & & & & & & & \(\Gamma_{1}\)\(\Gamma_{2}\)\(\Gamma_{6}\) \\ \hline
\end{tabular}
\caption{Optical selection rules at \(\Gamma\) point.} \label{tab:Gamma}
\end{center}
\end{table*}

{\centering {\subsection{\(\Lambda\) point}}}

\(\Lambda\) = (\( {\sqrt 3 \xi}/{2}, {\xi}/{2}\)). In \(R\bar{3}m\), \(G_{0}{(k)}\) = (\(\{E|0\}\), \(\{IC_{6}^{2}C_{x}|0\}\)) = \(C_{s}\), \(G/{G_{0}{(k)}}\) = (\(\{E|0\}\), \(\{C_{6}^{2}|0\}\), \(\{C_{6}^{4}|0\}\), \(\{I|0\}\), \(\{IC_{6}^{2}|0\}\), \(\{IC_{6}^{4}|0\}\)). In \(P6mm\), \(G_{0}{(k)}\) = (\(\{E|0\}\), \(\{IC_{3}C_{x}|0\}\)) = \(C_{s}\), \(G/{G_{0}{(k)}}\) = (\(\{E|0\}\), \(\{C_{6}|0\}\), \(\{C_{6}^{2}|0\}\), \(\{C_{6}^{3}|0\}\), \(\{C_{6}^{4}|0\}\), \(\{C_{6}^{5}|0\}\)). In \(P3m1\) \(G_{0}{(k)}\) = (\(\{E|0\}\), \(\{IC_{6}^{2}C_{x}|0\}\)) = \(C_{s}\), \(G/{G_{0}{(k)}}\) = (\(\{E|0\}\), \(\{C_{6}^{2}|0\}\), \(\{C_{6}^{4}|0\}\)).

As mentioned previously, \(-\Lambda\)\(\notin\)\(\bigstar \Lambda\) in \(P3m1\), while \(-\Lambda\)\(\in\)\(\bigstar \Lambda\) in \(R\bar{3}m\) and \(P6mm\). This has two effects: representations of \(\Lambda_{i}\) of \(R\bar{3}m\) and \(P6mm\) must decompose into pairs of \(\Lambda_{i}\) and \(\Lambda^{*}_{i}\) in \(P3m1\) (see Table~\ref{tab:restriction}), where \(\Lambda^{*}_{i}\) is irreducible representations of \(\bigstar\)\(-\Lambda\); zone center representations are contained in \(\Lambda_{i}\)\(\otimes\)\(\Lambda^{*}_{j}\) instead of in \(\Lambda_{i}\)\(\otimes\)\(\Lambda_{j}\) and \(\Lambda^{*}_{i}\)\(\otimes\)\(\Lambda^{*}_{j}\). This is also the case for \(Z\) point and \(R\) point.\\

\begin{table*}
\begin{center}
\begin{tabular}{cccccc} \hline
&& \multicolumn{2}{c}{\(R\bar{3}m\)}& \multicolumn{2}{c}{\(P6mm\)} \\ \hline
&& \(\Lambda_{1}\) & \(\Lambda_{2}\) &\(\Lambda_{1}\) &\(\Lambda_{2}\) \\ \hline
\multirow{2}{*}{\(R\bar{3}m\)}  & \(\Lambda_{1}\) &\(\Gamma_{1}\)\(\Gamma_{3}\)\(\Gamma_{4}\)\(\Gamma_{5}\)& \(\Gamma_{2}\)\(\Gamma_{3}\)\(\Gamma_{4}\)\(\Gamma_{6}\) &\(\Gamma_{1}\)\(\Gamma_{3}\)& \(\Gamma_{2}\)\(\Gamma_{3}\) \\
& \(\Lambda_{2}\) && \(\Gamma_{1}\),\(\Gamma_{3}\)\(\Gamma_{4}\)\(\Gamma_{5}\) &\(\Gamma_{2}\)\(\Gamma_{3}\)& \(\Gamma_{1}\)\(\Gamma_{3}\) \\ \hline
\multirow{2}{*}{\(P6mm\)}  & \(\Lambda_{1}\) &&&\(\Gamma_{1}\)\(\Gamma_{4}\)\(\Gamma_{5}\)\(\Gamma_{6}\) & \(\Gamma_{2}\)\(\Gamma_{3}\)\(\Gamma_{5}\)\(\Gamma_{6}\)\\
& \(\Lambda_{2}\) &&&&  \(\Gamma_{1}\)\(\Gamma_{4}\)\(\Gamma_{5}\)\(\Gamma_{6}\)\\ \hline
\end{tabular}
\caption{Optical selection rules at \(\Lambda\) point.} \label{tab:Lambda}
\end{center}
\end{table*}

{\centering {\subsection{\(\Sigma\) point}}}

\(\Sigma\) = (\(\xi, 0\)). In \(R\bar{3}m\), \(G_{0}{(k)}\) = (\(\{E|0\}\), \(\{C_{x}|0\}\)) = \(C_{2}\), \(G/{G_{0}{(k)}}\) = (\(\{E|0\}\), \(\{C_{6}^{2}|0\}\), \(\{C_{6}^{4}|0\}\), \(\{I|0\}\), \(\{IC_{6}^{2}|0\}\), \(\{IC_{6}^{4}|0\}\)). In \(P6mm\),  \(G_{0}{(k)}\) = (\(\{E|0\}\), \(\{IC_{6}^{3}C_{x}|0\}\)) = \(C_{s}\), \(G/{G_{0}{(k)}}\) = (\(\{E|0\}\), \(\{C_{6}|0\}\), \(\{C_{6}^{2}|0\}\), \(\{C_{6}^{3}|0\}\), \(\{C_{6}^{4}|0\}\), \(\{C_{6}^{5}|0\}\)). In \(P3m1\), \(G_{0}{(k)}\) = (\(\{E|0\}\)) = \(C_{1}\), \(G/{G_{0}{(k)}}\) = (\(\{E|0\}\), \(\{C_{6}^{2}|0\}\), \(\{C_{6}^{4}|0\}\), \(\{I|0\}\), \(\{IC_{6}^{2}|0\}\), \(\{IC_{6}^{4}|0\}\)). \\

\begin{table*}
\begin{center}
\begin{tabular}{cccccc} \hline
&& \multicolumn{2}{c}{\(R\bar{3}m\)}& \multicolumn{2}{c}{\(P6mm\)} \\ \hline
&& \(\Sigma_{1}\) & \(\Sigma_{2}\) &\(\Sigma_{1}\) &\(\Sigma_{2}\) \\ \hline
\multirow{2}{*}{\(R\bar{3}m\)}  & \(\Sigma_{1}\) &\(\Gamma_{1}\)\(\Gamma_{3}\)\(\Gamma_{4}\)\(\Gamma_{6}\)& \(\Gamma_{2}\)\(\Gamma_{3}\)\(\Gamma_{5}\)\(\Gamma_{6}\) &\(\Gamma_{1}\)\(\Gamma_{2}\)\(\Gamma_{3}\)& \(\Gamma_{1}\)\(\Gamma_{2}\)\(\Gamma_{3}\) \\
& \(\Sigma_{2}\) && \(\Gamma_{1}\)\(\Gamma_{3}\)\(\Gamma_{4}\)\(\Gamma_{5}\) &\(\Gamma_{1}\)\(\Gamma_{2}\)\(\Gamma_{3}\)& \(\Gamma_{1}\)\(\Gamma_{2}\)\(\Gamma_{3}\)\\ \hline
\multirow{2}{*}{\(P6mm\)}  & \(\Sigma_{1}\) &&&\(\Gamma_{1}\)\(\Gamma_{3}\)\(\Gamma_{5}\)\(\Gamma_{6}\) & \(\Gamma_{2}\)\(\Gamma_{4}\)\(\Gamma_{5}\)\(\Gamma_{6}\)\\
& \(\Sigma_{2}\) &&&&  \(\Gamma_{1}\)\(\Gamma_{3}\)\(\Gamma_{5}\)\(\Gamma_{6}\)\\ \hline
\end{tabular}
\caption{Optical selection rules at \(\Sigma\) point.} \label{tab:Sigma}
\end{center}
\end{table*}

{\centering {\subsection{\(M\) point}}}

\(M\) = (\({\pi}/{a}, {\pi}/{\sqrt{3}a}\)). In \(R\bar{3}m\), \(G_{0}{(k)}\) = (\(\{E|0\}\), \(\{IC_{6}^{2}C_{x}|0\}\)) = \(C_{s}\), \(G/{G_{0}{(k)}}\) = (\(\{E|0\}\), \(\{C_{6}^{2}|0\}\), \(\{C_{6}^{4}|0\}\), \(\{I|0\}\), \(\{IC_{6}^{2}|0\}\), \(\{IC_{6}^{4}|0\}\)). In \(P6mm\), \(G_{0}{(k)}\) = (\(\{E|0\}\), \(\{C_{6}^{3}|0\}\), \(\{IC_{6}C_{x}|0\}\), \(\{IC_{6}^{4}C_{x}|0\}\)) = \(C_{2v}\), \(G/{G_{0}{(k)}}\) = (\(\{E|0\}\), \(\{C_{6}|0\}\), \(\{C_{6}^{2}|0\}\)). In \(P3m1\), \(G_{0}{(k)}\) = (\(\{E|0\}\), \(\{IC_{6}^{4}C_{x}|0\}\)) = \(C_{s}\), \(G/{G_{0}{(k)}}\) =  (\(\{E|0\}\), \(\{C_{6}^{2}|0\}\), \(\{C_{6}^{4}|0\}\)). \\

\begin{table*}
\begin{center}
\begin{tabular}{cccccccc} \hline
&& \multicolumn{2}{c}{\(R\bar{3}m\)}&\multicolumn{4}{c}{\(P6mm\)}\\ \hline
&& \(M_{1}\) & \(M_{2}\) & \(M_{1}\) & \(M_{2}\) & \(M_{3}\) & \(M_{4}\) \\ \hline
\multirow{2}{*}{\(R\bar{3}m\)}&\(M_{1}\) & \(\Gamma_{1 }\)\(\Gamma_{3}\)\(\Gamma_{4}\)\(\Gamma_{5}\)& \(\Gamma_{2}\)\(\Gamma_{3}\)\(\Gamma_{4}\)\(\Gamma_{6}\)& \(\Gamma_{1}\)\(\Gamma_{3}\) & \(\Gamma_{2}\)\(\Gamma_{3}\) & \(\Gamma_{2}\)\(\Gamma_{3}\)& \(\Gamma_{1}\)\(\Gamma_{3}\) \\
&\(M_{2}\) && \(\Gamma_{1}\)\(\Gamma_{3}\)\(\Gamma_{4}\)\(\Gamma_{5}\)& \(\Gamma_{2}\)\(\Gamma_{3}\) &\(\Gamma_{1}\)\(\Gamma_{3}\)&\(\Gamma_{1}\)\(\Gamma_{3}\)&\(\Gamma_{2}\)\(\Gamma_{3}\)\\ \hline
\multirow{4}{*}{\(P6mm\)}&\(M_{1}\) &&&\(\Gamma_{1}\)\(\Gamma_{6}\) & \(\Gamma_{2}\)\(\Gamma_{6}\) & \(\Gamma_{3}\)\(\Gamma_{5}\) & \(\Gamma_{4}\)\(\Gamma_{5}\)\\
&\(M_{2}\)&&&& \(\Gamma_{1}\)\(\Gamma_{6}\) & \(\Gamma_{4}\)\(\Gamma_{5}\) & \(\Gamma_{3}\)\(\Gamma_{5}\) \\
&\(M_{3}\)&&&&& \(\Gamma_{1}\)\(\Gamma_{6}\) & \(\Gamma_{2}\)\(\Gamma_{6}\) \\
&\(M_{4}\)&&&&&& \(\Gamma_{1}\)\(\Gamma_{6}\) \\ \hline
\end{tabular}
\caption{Optical selection rules at \(M\) point.} \label{tab:M}
\end{center}
\end{table*}

{\centering {\subsection{\(K\) point}}}

\(K\) = (\({4\pi}/{3a}, 0\)). In \(R\bar{3}m\), \(G_{0}{(k)}\) = (\(\{E|0\}\), \(\{C_{x}|0\}\)) = \(C_{2}\), \(G/{G_{0}{(k)}}\) = (\(\{E|0\}\), \(\{C_{6}^{2}|0\}\), \(\{C_{6}^{4}|0\}\), \(\{I|0\}\), \(\{IC_{6}^{2}|0\}\), \(\{IC_{6}^{4}|0\}\)). In \(P6mm\), \(G_{0}{(k)}\) = (\(\{E|0\}\), \(\{C_{6}^{2}|0\}\), \(\{C_{6}^{4}|0\}\), \(\{IC_{6}^{1}C_{x}|0\}\), \(\{IC_{6}^{3}C_{x}|0\}\), \(\{IC_{6}^{5}C_{x}|0\}\)) = \(C_{3v}\), \(G/{G_{0}{(k)}}\) = (\(\{E|0\}\), \(\{C_{6}^{1}|0\}\)). In \(P3m1\), \(G_{0}{(k)}\) = (\(\{E|0\}\), \(\{C_{6}^{2}|0\}\), \(\{C_{6}^{4}|0\}\)) = \(C_{3}\), \(G/{G_{0}{(k)}}\) = (\(\{E|0\}\), \(\{IC_{x}|0\}\)). 

For \(K\) point of \(P3m1\), \(K_{2}\) and \(K_{3}\) are connected by the time reversal operation and any physical representation must be decomposed into equal numbers of \(K_{2}\) and \(K_{3}\) representations, as shown in Table~\ref{tab:restriction}. Zone center representations are obtained by doing the direct products of \(K_{1}\)\(\otimes\)\(K_{1}\), \(K_{1}\)\(\otimes\)(\(K_{2}\)+\(K_{3}\)) and (\(K_{2}\)+\(K_{3}\))\(\otimes\)(\(K_{2}\)+\(K_{3}\)).\\

\begin{table*}
\begin{center}
\begin{tabular}{ccccccc} \hline
&& \multicolumn{2}{c}{\(R\bar{3}m\)}&\multicolumn{3}{c}{\(P6mm\)}\\ \hline
&& \(K_{1}\) & \(K_{2}\) & \(K_{1}\) & \(K_{2}\) & \(K_{3}\)  \\ \hline
\multirow{2}{*}{\(R\bar{3}m\)}&\(K_{1}\) & \(\Gamma_{1 }\)\(\Gamma_{3}\)\(\Gamma_{4}\)\(\Gamma_{6}\)& \(\Gamma_{2}\)\(\Gamma_{3}\)\(\Gamma_{5}\)\(\Gamma_{6}\)& \(\Gamma_{1}\)\(\Gamma_{2}\)\(\Gamma_{3}\) & \(\Gamma_{1}\)\(\Gamma_{2}\)\(\Gamma_{3}\) & \(\Gamma_{1}\)\(\Gamma_{2}\)\(\Gamma_{3}\)\\
&\(K_{2}\) && \(\Gamma_{1}\)\(\Gamma_{3}\)\(\Gamma_{4}\)\(\Gamma_{6}\)& \(\Gamma_{1}\)\(\Gamma_{2}\)\(\Gamma_{3}\) &\(\Gamma_{1}\)\(\Gamma_{2}\)\(\Gamma_{3}\)&\(\Gamma_{1}\)\(\Gamma_{2}\)\(\Gamma_{3}\) \\ \hline
\multirow{3}{*}{\(P6mm\)}&\(K_{1}\) &&&\(\Gamma_{1}\)\(\Gamma_{3}\) & \(\Gamma_{2}\)\(\Gamma_{4}\) & \(\Gamma_{5}\)\(\Gamma_{6}\)\\
&\(K_{2}\)&&&& \(\Gamma_{1}\)\(\Gamma_{3}\) & \(\Gamma_{5}\)\(\Gamma_{6}\) \\
&\(K_{3}\)&&&&& \(\Gamma_{1}\)\(\Gamma_{2}\)\(\Gamma_{3}\)\(\Gamma_{4}\)\(\Gamma_{5}\)\(\Gamma_{6}\) \\ \hline
\end{tabular}
\caption{Optical selection rules at \(K\) point.} \label{tab:K}
\end{center}
\end{table*}

{\centering {\subsection{\(Z\) point}}}

\(Z\) = (\(2\xi, {8\pi}/{3a}-2\xi\)). In \(R\bar{3}m\), \(G_{0}{(k)}\) = (\(\{E|0\}\)) = \(C_{1}\), \(G/{G_{0}{(k)}}\) = (\(\{E|0\}\), \(\{C_{6}^{2}|0\}\), \(\{C_{6}^{4}|0\}\), \(\{C_{x}|0\}\), \(\{C_{6}^{2}C_{x}|0\}\), \(\{C_{6}^{4}C_{x}|0\}\), \(\{I|0\}\), \(\{IC_{6}^{2}|0\}\), \(\{IC_{6}^{4}|0\}\), \(\{IC_{x}|0\}\), \(\{IC_{6}^{2}C_{x}|0\}\), \(\{IC_{6}^{4}C_{x}|0\}\)). In \(P6mm\), \(G_{0}{(k)}\) = (\(\{E|0\}\), \(\{IC_{6}^{1}C_{x}|0\}\)) = \(C_{s}\), \(G/{G_{0}{(k)}}\) = (\(\{E|0\}\), \(\{C_{6}|0\}\), \(\{C_{6}^{2}|0\}\), \(\{C_{6}^{3}|0\}\), \(\{C_{6}^{4}|0\}\), \(\{C_{6}^{5}|0\}\)). In \(P3m1\), \(G_{0}{(k)}\) = (\(\{E|0\}\)) = \(C_{1}\), \(G/{G_{0}{(k)}}\) = (\(\{E|0\}\), \(\{C_{6}^{2}|0\}\), \(\{C_{6}^{4}|0\}\), \(\{IC_{x}|0\}\), \(\{IC_{6}^{2}C_{x}|0\}\), \(\{IC_{6}^{4}C_{x}|0\}\)). \\

\begin{table*}
\begin{center}
\begin{tabular}{ccccc} \hline
&& \(R\bar{3}m\) & \multicolumn{2}{c}{\(P6mm\)} \\ \hline
&& \(Z_{1}\) & \(Z_{1}\) & \(Z_{2}\) \\ \hline
\(R\bar{3}m\) & \(Z_{1}\) & \(\Gamma_{1}\)\(\Gamma_{2}\)\(\Gamma_{3}\)\(\Gamma_{4}\)\(\Gamma_{5}\)\(\Gamma_{6}\) & \(\Gamma_{1}\)\(\Gamma_{2}\)\(\Gamma_{3}\) & \(\Gamma_{1}\)\(\Gamma_{2}\)\(\Gamma_{3}\) \\ \hline
\multirow{2}{*}{\(P6mm\)} &\(Z_{1}\) && \(\Gamma_{1}\)\(\Gamma_{2}\)\(\Gamma_{3}\)\(\Gamma_{4}\)\(\Gamma_{5}\)\(\Gamma_{6}\) & \(\Gamma_{1}\)\(\Gamma_{2}\)\(\Gamma_{3}\)\(\Gamma_{4}\)\(\Gamma_{5}\)\(\Gamma_{6}\) \\
&\(Z_{2}\) &&& \(\Gamma_{1}\)\(\Gamma_{2}\)\(\Gamma_{3}\)\(\Gamma_{4}\)\(\Gamma_{5}\)\(\Gamma_{6}\)\\ \hline
\end{tabular}
\caption{Optical selection rules at \(Z\) point.} \label{tab:Z}
\end{center}
\end{table*}

{\centering {\subsection{\(R\) point}}}

\(R\) = (\(\xi,\zeta\)) is the arbitrary point in the Brillouin zone without any symmetry. In \(R\bar{3}m\), \(G_{0}{(k)}\) = (\(\{E|0\}\)) = \(C_{1}\), \(G/{G_{0}{(k)}}\) = (\(\{E|0\}\), \(\{C_{6}^{2}|0\}\), \(\{C_{6}^{4}|0\}\), \(\{C_{x}|0\}\), \(\{C_{6}^{2}C_{x}|0\}\), \(\{C_{6}^{4}C_{x}|0\}\), \(\{I|0\}\), \(\{IC_{6}^{2}|0\}\), \(\{IC_{6}^{4}|0\}\), \(\{IC_{x}|0\}\), \(\{IC_{6}^{2}C_{x}|0\}\), \(\{IC_{6}^{4}C_{x}|0\}\)). In \(P6mm\), \(G_{0}{(k)}\) = (\(\{E|0\}\)) = \(C_{1}\), \(G/{G_{0}{(k)}}\) = (\(\{E|0\}\), \(\{C_{6}|0\}\), \(\{C_{6}^{2}|0\}\), \(\{C_{6}^{3}|0\}\), \(\{C_{6}^{4}|0\}\), \(\{C_{6}^{5}|0\}\),\(\{IC_{x}|0\}\), \(\{IC_{6}^{1}C_{x}|0\}\), \(\{IC_{6}^{2}C_{x}|0\}\), \(\{IC_{6}^{3}C_{x}|0\}\), \(\{IC_{6}^{4}C_{x}|0\}\), \(\{IC_{6}^{5}C_{x}|0\}\)). In \(P3m1\), \(G_{0}{(k)}\) = (\(\{E|0\}\)) = \(C_{1}\), \(G/{G_{0}{(k)}}\) = (\(\{E|0\}\), \(\{C_{6}^{2}|0\}\), \(\{C_{6}^{4}|0\}\), \(\{IC_{x}|0\}\), \(\{IC_{6}^{2}C_{x}|0\}\), \(\{IC_{6}^{4}C_{x}|0\}\)).

\begin{table*}
\begin{center}
\begin{tabular}{cccc} \hline
&& \(R\bar{3}m\) & \(P6mm\) \\ \hline
&& \(R_{1}\) & \(R_{1}\) \\ \hline
\(R\bar{3}m\) & \(R_{1}\) & \(\Gamma_{1}\)\(\Gamma_{2}\)\(\Gamma_{3}\)\(\Gamma_{4}\)\(\Gamma_{5}\)\(\Gamma_{6}\) & \(\Gamma_{1}\)\(\Gamma_{2}\)\(\Gamma_{3}\) \\ \hline
\(P6mm\) &\(R_{1}\) && \(\Gamma_{1}\)\(\Gamma_{2}\)\(\Gamma_{3}\)\(\Gamma_{4}\)\(\Gamma_{5}\)\(\Gamma_{6}\) \\ \hline
\end{tabular}
\caption{Optical selection rules at \(R\) point.} \label{tab:R}
\end{center}
\end{table*}

While doing the direct products for \(R\bar{3}m\), \(P6mm\) and \(P3m1\) at arbitrary point \(R\), a general rule about zone center representations in the direct product is discovered: the direct product of any physical representation at arbitrary point \(R\) of the Brillouin zone with itself constains zone center representations, the number of appearence of the zone center representations equals the dimensionality of the representation. For example, without calculation, at arbitrary point \(R\) in \(D_{4h}^{7}\), \(\bigstar{R_{1}} \otimes \bigstar{R_{1}}  = \Gamma_{1}^{+} + \Gamma_{2}^{+} + \Gamma_{3}^{+} + \Gamma_{4}^{+} + 2\Gamma_{5}^{+} + \Gamma_{1}^{-} + \Gamma_{2}^{-} + \Gamma_{3}^{-} + \Gamma_{4}^{-} + 2\Gamma_{5}^{-}\), plus many terms away from zone center. The proof is given elsewhere.\\

Tables~\ref{tab:Gamma}--\ref{tab:R} cover the entire Brillouin zone of the surface space group \(P6mm\). In tables~\ref{tab:Gamma}--\ref{tab:R}, a representation in the direct products is only written down once even if it appears many times because one is more concerned about whether or not a transition can happen instead of its number of appearances in the direct product. Careful inspection of tables~\ref{tab:Gamma}--\ref{tab:R} shows that for optical transitions between bulk state and surface state, polarization effects only exist at \(\Gamma\), \(\Lambda\) and \(M\) points, where light polarized in different directions participates in different transitions. This is in contrast with \(\Sigma\), \(K\), \(Z\) and \(R\) points where \(x, y, z\) is contained in all direct products therefore light polarized in different directions participate in all transitions. The coupling strength is, of course, polarization dependent.

To demonstrate the use of the tables, consider \(M\) point (Table~\ref{tab:M}). \(p_{x}\), \(p_{y}\), \(p_{z}\) atomic orbits transform as \(M_{3}\), \(M_{4}\), \(M_{1}\) of space group \(P6mm\) and \(M_{2}\), \(M_{2}\), \(M_{1}\) of space group \(R\bar{3}m\). While photons polarized in the \((x, y)\) direction (\(\Gamma_{3}\)) participate in all bulk to surface transitions, \(z\) direction polarized photons (\(\Gamma_{1}\)) only participate in transitions from bulk \(M_{1}\) to surface \(M_{1}\), \(M_{4}\) and from bulk \(M_{2}\) to surface \(M_{2}\), \(M_{3}\).\\

{\centering{\section{Application to photoemission process}}}

Electron spin polarization (ESP) of photoelectrons of TIs will now be calculated to demonstrate the use of the tables. ESP measuments have the advantage that it depends entirely and only on the symmetry properties of the wave functions involved and is independent of the form of the crystal potential. ESP of photoelectrons has been calculated for \ce{GaAs} and the results agree well with experiments~\cite{wohlecke}. Instead of the total photoelectron intensity, ESP measures the degree of spin polarization, which is defined as \[P = \frac{I\uparrow - I\downarrow}{I\uparrow + I\downarrow},\] where \(I\uparrow\) (\(I\downarrow\)) denotes the intensity of spin up (down) electrons. ESP at \(\Gamma\) point is calculated for surface to bulk and bulk to surface transitions. Such transitions are the first step of the photoemission process in the ``three step model"~\cite{Spicer}.

\begin{table*}
\begin{centering}
\begin{tabular}{ccccccccc} \hline
&\(\Gamma_{7}^{1}\)& \(\Gamma_{7}^{2}\)& \(\Gamma_{7}^{3}\)& \(\Gamma_{8}^{4}\)& \(\Gamma_{7}^{5}\)& \(\Gamma_{9}^{5}\)& \(\Gamma_{8}^{6}\)& \(\Gamma_{9}^{6}\) \\ \hline
\(\Gamma_{4+}^{1+}\)&0 &0& 0& 0& 1& -1& 1& 1 \\
\(\Gamma_{4+}^{2+}\)&0 &0& 0& 0& 1& -1& 1& 1 \\
\(\Gamma_{4+}^{3+}\)&-1 &-1& 0& 1& -1& -1& 0& 1 \\
\(\Gamma_{5+}^{3+} \oplus \Gamma_{5+}^{3+}\)&1 &1& -1& 0& 1& 1& -1& 0 \\
\(\Gamma_{4-}^{1-}\)&0 &0& 0& 0& 1& -1& 1& 1 \\
\(\Gamma_{4-}^{2-}\)&0 &0& 0& 0& 1& -1& 1& 1 \\
\(\Gamma_{4-}^{3-}\)&-1 &-1& 0& 1& -1& -1& 0& 1 \\
\(\Gamma_{5-}^{3-} \oplus \Gamma_{5-}^{3-}\) &1 &1& -1& 0& 1& 1& -1& 0 \\ \hline
\end{tabular}
\caption{Electron spin polarization of photoelectrons excited from bulk band to surface band, with right circularly polarized light. Bulk electronic states are on the left and surface electronic states are at the top.} \label{tab:bulktosurface}
\end{centering}
\end{table*}

Taking into account of electron spin, the electronic wave functions are described by double groups. In point group \(D_{3d}\), the double group representations are \(\Gamma_{4\pm}\), \(\Gamma_{5\pm}\) and \(\Gamma_{6\pm}\). In the presence of the time reversal symmetry \(\hat \theta\), \(\Gamma_{5+}\) and \(\Gamma_{6+}\) stick together and \(\Gamma_{5-}\) and \(\Gamma_{6-}\) stick together. In point group \(C_{6v}\), the double group representations are \(\Gamma_{7}\), \(\Gamma_{8}\) and \(\Gamma_{9}\). In point group \(C_{3v}\), the double group representations are  \(\Gamma_{4}\), \(\Gamma_{5}\) and \(\Gamma_{6}\), where \(\hat \theta\) forces \(\Gamma_{5}\) and \(\Gamma_{6}\) together. The restrictions of double group representations from \(R\bar{3}m\) and \(P6mm\) to \(P3m1\) at \(\Gamma\) point are: \[\begin{array}{ccccccc}
R\bar{3}m&\rightarrow& P3m1&&P6mm &\rightarrow&P3m1 \\
\Gamma_{4\pm}&\rightarrow& \Gamma_{4}&&\Gamma_{7} &\rightarrow&\Gamma_{4} \\
\Gamma_{5+} \oplus \Gamma_{6+}&\rightarrow& \Gamma_{5} \oplus \Gamma_{6}&&\Gamma_{8}&\rightarrow&\Gamma_{4} \\
\Gamma_{5-} \oplus \Gamma_{6-}&\rightarrow& \Gamma_{5} \oplus \Gamma_{6}&&\Gamma_{9} &\rightarrow&\Gamma_{5} \oplus \Gamma_{6}
\end{array} \]
In ESP calculations double group representations are written down in the form of \(\Gamma_{i}^{j}\), which means that the wave function transforms as \(\Gamma_{i}\) double group and the spatial part belongs to \(\Gamma_{j}\) single group representation. The results of ESP calculations are tabulated in table~\ref{tab:bulktosurface} for bulk to surface transitions and in table~\ref{tab:surfacetobulk} for surface to bulk transitions. The results are calculated for right circularly polarized light (\(X - {\rm i} Y\)) at the center of the Brillouin zone. It is known that hybridization effect may lower the magnitude of ESP values~\cite{Meier}.\\

\begin{table*}
\begin{centering}
\begin{tabular}{ccccccccc} \hline
&\(\Gamma_{4+}^{1+}\)& \(\Gamma_{4+}^{2+}\)& \(\Gamma_{4+}^{3+}\)& \(\Gamma_{5+}^{3+} \oplus \Gamma_{5+}^{3+}\)& \(\Gamma_{4-}^{1-}\)& \(\Gamma_{4-}^{2-}\)& \(\Gamma_{4-}^{3-}\)& \(\Gamma_{5-}^{3-} \oplus \Gamma_{5-}^{3-}\) \\ \hline
\(\Gamma_{7}^{1}\)&0 &0& 1& -1& 0& 0& 1& -1 \\
\(\Gamma_{7}^{2}\)&0 &0& 1& -1& 0& 0& 1& -1 \\
\(\Gamma_{7}^{3}\)&0 &0& 1& -1& 0& 0& 1& -1 \\
\(\Gamma_{8}^{4}\)&0 &0& 1& -1& 0& 0& 1& -1 \\
\(\Gamma_{7}^{5}\)&-1 &-1& 0& 1& -1& -1& 0& 1 \\
\(\Gamma_{9}^{5}\)&1 &1& -1& 0& 1& 1& -1& 0 \\
\(\Gamma_{8}^{6}\)&-1 &-1& 0& 1& -1& -1& 0& 1 \\
\(\Gamma_{9}^{6}\)&1 &1& -1& 0& 1& 1& -1& 0 \\ \hline
\end{tabular}
\caption{Electron spin polarization of photoelectrons excited from surface band to bulk band, with right circularly polarized light. Surface electronic states are on the left and bulk electronic states are at the top.} \label{tab:surfacetobulk}
\end{centering}
\end{table*}

{\centering {\section{CONCLUSIONS}}}

Despite all the possible applications of our analysis, the use of \(P6mm\) wall paper group must be justified. Surface reconstructions are important for crystal surfaces. It has been well studied on the \ce{Si} (111) surface and several reconstruction patterns were found~\cite{Hanemann, Pandey}. To achieve lower energy, reconstruction tends to reduce the symmetry of the surfaces. The atoms in \ce{V2VI3} are stacked in the order of \(A\), \(B\), \(C\), \(A\), \(\cdots\). There is a mismatch between \(A\), \(B\) and \(C\) sites. Seen from the surface layer, the second and the third layers (see Fig.~\ref{fig:structure}\(b\)) have only three fold symmetry. Interactions between the surface layer and inner layers will be three fold like. As a result, the surface structure of TIs could lose the \(C_{6v}\) symmetry. However, to our knownledge, no surface reconstructions was reported in these systems~\cite{ZhangAPL, Urazhdin}. The perfect six fold symmetry of the Fermi surface is clearly demonstrated in ARPES~\cite{Chen, Bianchi} and STM~\cite{ZhangPRL, Kim, Alpichshev} experiments. Of perticular interest is reference~\cite{Chen}, where both the six fold symmetry of surface states and the three fold symmetry of bulk states are resolved simultaneously. These observations indicate that surface structure of TIs indeed belongs to \(P6mm\) (which has been previously noticed by Zhu {\em et al.}~\cite{Zhu}) and surface electronic wave functions can be characterized in \(P6mm\). This paper is not intended to discuss the stability of the surface states (that are topologically protected against non-magnetic impurities), but rather only focus on their symmetry properties which govern selections rules.

The actual degree of distortion of surface state wave functions in the \(z\) direction due to interactions with inner layers requires numerical calculations and is beyond the scope of this paper. Wave function distortion in the \(z\) direction, however, does not invalidate group theoretical calculations because \(z\) direction distorted wave functions are still basis functions of wallpaper groups. Distortions of wave functions only change the magnitude of coupling constants between surface states and bulk states.\\

In summary, we performed group theoretical study on topological insulators over the entire Brillouin zone. A new method is used to handle transitions between bulk states and surface states. It is shown that polarization effects only exist along the \(\Gamma\)\(M\) line. The results of electon spin polarization calculations of photoelectrons at Brillouin zone center are tabulated. Our work will be helpful for future optical studies on topological insulators. It's our next objective to explain the existing optical experiments within the framework of present group theoretical analysis, perticularly for transitions between bulk and surface. Many authors may be considering the problem only along certain Brillouin zone directions such as \(\Gamma\)\(K\) or \(\Gamma\)\(M\) while the analysis in the entire Brillouin zone is more appropriate due to spin momentum locking: the electrons are not in a simple spin up or spin down state.\\

\end{document}